\documentclass[onecolumn,10pt,showpacs]{revtex4}
\usepackage{graphicx}
\usepackage{epsfig}
\usepackage{bm}
\usepackage{amsfonts}

\input epsf

\begin{document}

\title{\Large Accretion of New Variable Modified Chaplygin Gas and
Generalized Cosmic Chaplygin Gas onto Schwarzschild and Kerr-Newman Black holes}

\author{\bf Jhumpa Bhadra$^1$\footnote{bhadra.jhumpa@gmail.com}
and Ujjal Debnath$^1$\footnote{ujjaldebnath@yahoo.com ,
ujjal@iucaa.ernet.in}}

\affiliation{$^1$ Department of Mathematics,
Bengal Engineering and Science University, Shibpur, Howrah-711
103, India.\\}

\date{\today}

\begin{abstract}
In this work, we have studied accretion of the dark energies like
new variable modified Chaplygin gas (NVMCG) and generalized cosmic
Chaplygin gas (GCCG) onto Schwarzschild and Kerr-Newman Black
holes. We find the expression of the critical four velocity
component which gradually decreases for the fluid flow towards the
Schwarzschild as well as Kerr-Newman Black hole. We also find the
expression for change of masses of the black hole in both cases.
For the Kerr-Newman black hole which is rotating and charged we
calculate the specific angular momentum and total angular
momentum. We showed that in both cases due to accretion of the
dark energy mass of the black hole increases and angular momentum
increases in case of Kerr-Newman black hole.
\end{abstract}

\pacs{04.70.Bw, 04.70.Dy, 98.80.Cq}

\maketitle

\sloppy \tableofcontents

\section{Introduction}
Different observational data together with observations of
supernovae of type Ia
\cite{{Bachall},{Perlmutter1},{Perlmutter2},{Riess}}, WMAP
\cite{Bennett}, Chandra X-ray Observatory \cite{Allen}  strongly
indicate that our universe is undergoing an accelerating phase.
Nonbaryonic matter recognized as dark energy  having negative
pressure and violate the strong energy
\cite{{Sahni1},{Peebles0},{Padmanabhan}} condition may explain
this accelerated expansion. There are various candidates to play
the role of the dark energy which is the dominant part of the
universe. The dark energy candidates are cosmological constant
\cite{Sami}, quintessence, K-essence \cite{Armendariz-Picon},
Chaplygin gas \cite{Kamenshchik}, its modification known as
modified Chaplygin gas (MCG) \cite{Debnath}, tachyonic field
\cite{Sen}, DBI-essence \cite{Martin} etc. The Chaplygin Gas (EoS
$p=-\frac{B}{\rho}$, $B>0$) \cite{Kamenshchik} acts as
pressureless fluid for small value of the scale factor and tends
to accelerated expansion for large value of scale factor.
Generalization of Chaplygin gas model known as {\it Generalized
Chaplygin gas} which satisfies $p=-\frac{B}{\rho^{\alpha}}$, $0
\leq \alpha \leq 1$ \cite{{Gorini},{Alam},{Bento}}. This model
also modified to {\it Modified Chaplygin Gas} (MCG) having the Eos
$p=A \rho-\frac{B}{\rho^\alpha}$, $0 \leq \alpha \leq 1$, $A>0,
B>0$ \cite{{Benaoum},{Sahni},{Debnath}}. That illustrates a
radiation era $(A =1/3)$ while the scale factor  is vanishingly
small and $\Lambda$CDM model for infinitely large scale factor.
Further Guo and Zhang \cite{Guo} established {\it Variable
Chaplygin Gas} with Eos is $p=-\frac{B}{\rho}$, where $B=B(R)$,
$R$ is the scale factor and $B$ is a positive function of the
scale factor. Subsequently Debnath \cite{Debnath1} provided {\it
Variable modified Chaplygin Gas} with Eos is
$p=A\rho-\frac{B(R)}{\rho^\alpha}$ for the accelerating phase of
the universe. The another candidate of dark energy was introduced
by Chakraborty et al \cite{UDebnath}, known as {\it New Variable
modified Chaplygin Gas} (NVMCG) which follows the equation
$p=A(R)-\frac{B(R)}{\rho^\alpha}$, $0 \leq \alpha \leq 1$ which
gives interesting physical significance. In 2003 P.F.
Gonz$\acute{a}$lez-Diaz \cite{Diaz} gives the idea of another form
of dark energy to the consequence of accelerating phase of
universe namely {\it Generalized Cosmic Chaplygin Gas} (GCCG),
This model is stable and free from unphysical behaviour even when
the vacuum fluid satisfies the phantom energy condition.\\

A cosmological property in which there is an infinite expansion in
scale factor in a finite time termed as `Big Rip' \cite{Caldwell}.
In the phantom cosmology, big rip is a kind of future singularity
in which the energy density of phantom energy ($\rho+p<0$) will
become infinite in a finite time. To realize the Big Rip scenario
the condition $\rho+p<0$ alone is not sufficient \cite{McInnes}.
Distinct data on supernovas showed that the presence of phantom
energy with $-1.2 < w < -1$ in the Universe is highly likely
\cite{Alam2}. The accretion of phantom dark energy onto a
Schwarzschild black hole was first modelled by Babichev et al
\cite{Babichev}. They  established that black hole mass will
gradually decrease due to strong negative pressure of phantom
energy and finally all the masses tend to zero near the big rip
where it will disappear. Accretion of phantom like modified
variable Chaplygin gas onto Schwarzschild black hole was studied
by Jamil \cite{Jamil} who showed that mass of the black hole will
decrease  when accreting fluid violates the dominant energy
condition and otherwise will increase. Also the accretion of dark
energy with EoS $p=w \rho$ onto the Kerr-Newman black hole was
studied by Madrid et al \cite{Pedro} and they obtained that if
$w>-1$, mass and angular momentum increase. Mass of the black hole
grows up unboundedly whereas the angular momentum increases up to a given level.\\

In the present work, we have studied accretion of dark energy
namely new variable modified Chaplygin gas (NVMCG) and generalized
cosmic Chaplygin gas (GCCG) onto Schwarzschild as well as most
generalized Kerr-Newman black holes. For natures of black hole
mass function with angular momentum have been analyzed when NVMCG
and GCCG like dark energies accrete upon black holes.\\

\section{Accretion of dark energy onto Schwarzschild black hole}

Let us consider a spherically symmetrical accretion of the dark
energy onto the black hole. We consider a Schwarzschild black hole
(static) of mass $M$ which is gravitationally isolated (in
geometrical units, $G=1=c$) \cite{Babichev, Jamil} described by
the line element
\begin{eqnarray}
ds^2=\left(1-\frac{2M}{r}\right) dt^2-\left(1-\frac{2M}{r}\right)^{-1} dr^2-r^2\left(d\theta^2+\sin^2 \theta d\phi^2\right)
\end{eqnarray}

where, $r$ being the radial coordinate. Energy momentum-tensor for
the DE, considering in the form of perfect fluid having the EoS
$p=p(\rho)$, is

\begin{eqnarray}
T_{\mu\nu}=(\rho+p)u_\mu u_\nu-p g_{\mu\nu}
\end{eqnarray}

where $\rho$, $p$  are the density and pressure of the dark energy
respectively and $u^\mu=\frac{dx^\mu}{ds}$ is the fluid
4-velocity satisfying $u^\mu u_\mu=1$. We assume that the in-falling
dark energy fluid does not disturb the spherical symmetry of the black hole.\\

The relativistic Bernoulli's equation after the time component of
the energy-momentum conservation law $T^{\mu\nu}_{;\nu}=0$ provide
the first integral of motion  for stationary, spherically
symmetric accretion onto BH which yields

\begin{eqnarray}
(\rho+p)\left(1-\frac{2}{x}+u^2\right)^{\frac{1}{2}} x^2 u=C_1
\end{eqnarray}

where $x=\frac{r}{M}$ and $u=\frac{dr}{ds}$ is the radial
component of the velocity four vector and $C_1$ being the
integrating constant. In the case of fluid flow directed
towards the black hole, we must have $u<0$.\\

Moreover, the second integration of motion is obtained from $u_\mu
T^{\mu \nu}_{;\nu}=0$, which gives

\begin{eqnarray}
ux^2 \exp{\left[\int^{\rho}_{\rho_\infty}\frac{d \rho}{\rho+p(\rho)}\right]}=-A
\end{eqnarray}

where, $A (>0)$ is the integration constant, $\rho_{\infty}$ is
the dark energy density at infinity. Further value of the constant
$A$
is evaluated for different DE model.\\

Using the equations (3) and (4) we get
\begin{eqnarray}
(\rho+p)\sqrt{1-\frac{2}{x}+u^2}\exp{\left[-\int^{\rho}_{\rho_\infty}\frac{d \rho}{\rho+p(\rho)}\right]}=C_2
\end{eqnarray}

where $C_2=-C_1/A=\rho_{\infty}+p(\rho_{\infty})$.\\

If $n$ be the concentration of dark energy which satisfies the following equations

\begin{eqnarray}
\frac{n(\rho)}{n_\infty}=\exp\left[{\int_{\rho_\infty}^{\rho}}\frac{d\rho}{\rho+p(\rho)}\right]
\end{eqnarray}

where $n_\infty=n(\rho_\infty)$ being the concentration of the dark energy at infinity.\\

The constant value $A$ can be determined by finding the critical
point of the accretion using \cite{Michel} then,

\begin{eqnarray}
u_*^2=\frac{1}{2 x_*}~~~~~~~~~~~~~~c_s^2(\rho_*)=\frac{u_*^2}{1-3u_*^2}
\end{eqnarray}

where $c_s=\sqrt{\frac{\partial p}{\partial \rho}}$ ~is the usual
speed of sound, $u_*$ is the critical four velocity component and
$\rho_{*}$ is the density at the critical point $x_{*}$. One may
noted that for $c_{s}^{2}>0$ or $c_{s}^{2}<1$, no critical point
exists outside the black hole (i.e., $x_{*}>2$). Using (5) and
(7), we get the following relation

\begin{eqnarray}
\frac{\rho_*+p(\rho_*)}{\rho_\infty+p(\rho_\infty)}=\sqrt{1+3 c_s^2(\rho_*)}\exp\left[{\int_{\rho_\infty}^{\rho_*}}\frac{d\rho}{\rho+p(\rho)}\right]
\end{eqnarray}

The rate of change of mass $\dot{M}$ of the black hole is computed
by integrating the flux of the dark energy over the entire horizon
of the black hole i.e., $\dot{M}=\oint T_{t}^{r} dS$, where
$T_{t}^{r}$ represents the radial component of the energy momentum densities
and the surface element of the black hole horizon $dS=\sqrt{-g} d\theta d\phi$ \cite{Babichev}. \\

Using the above equations we obtain the rate of change of mass as

\begin{eqnarray}
\dot{M}=4 \pi A M^2(\rho+p)
\end{eqnarray}

Since the Schwarszchild black hole is static, so the mass of the
black hole depends on $r$ only. When some fluid accretes outside
Schwarszchild the black hole, the mass function $M$ of the black
hole is considered as a dynamical mass function and hence it
should be a function of time also. So $\dot{M}$ of the equation
(9) is time dependent and the increasing or decreasing of the
black hole mass $M$ sensitively depends on the nature of the
fluid which accretes upon the black hole.\\

At the black hole horizon ($r=2M$ i.e., $x=2$) the relation
between four velocity $u_H=u(\rho_H)$ and the energy density
$\rho_H$ at the black hole event horizon is given by

\begin{eqnarray}
\frac{A}{4}\frac{\rho_H+p(\rho_H)}{\rho_\infty+p(\rho_\infty)}=\frac{A^2}{16 u_H^2}
=\exp\left[2{\int_{\rho_\infty}^{\rho_H}}\frac{d\rho}{\rho+p(\rho)}\right]
\end{eqnarray}

and velocity four component at the horizon of black hole is given by
\begin{eqnarray}
u_H=-\frac{A}{4}\exp\left[{\int_{\rho_\infty}^{\rho_H}}\frac{d\rho}{\rho+p(\rho)}\right]
\end{eqnarray}

\subsection{Model I: New Variable Modified Chaplygin Gas as dark energy model}

We consider the background spacetime is spatially flat represented
by the homogeneous and isotropic FRW model of the universe which
is given by

$$
ds^{2}=-dt^{2}+R^{2}(t)\left[dr^{2}+r^{2}(d\theta^{2}+sin^{2}\theta
d\phi^{2}) \right]
$$

where $R(t)$ is the scale factor. We assume the universe is filled
with New Variable Modified Chaplygin Gas (NVMCG) and the EoS is
\cite{UDebnath} given by

\begin{eqnarray}
p=A'(R)\rho-\frac{B(R)}{\rho^\alpha} ~~~~ \mbox{with } 0\leq \alpha \leq 1
\end{eqnarray}

where $A'(R)$, and $B(R)$ are function of the scale factor $R$.
For a particular choice $A'(R)=A_0 R^{-n}$ and $B(R)=B_0 R^{-m}$
with $A_0$, $B_0$, $m$, $n$ are positive constants. For $n=m=0$,
this model reduces to modified Chaplygin Gas, and for $n=0$ the
model reduces to the variable modified Chaplygin gas model.\\

The Einstein's equations for FRW universe are (choosing $G=c=1$)

\begin{eqnarray} H^2 = \frac{8 \pi}{3} \rho \end{eqnarray}

\begin{eqnarray} \dot{H}=-\frac{8 \pi }{2}\left(p + \rho \right)\end{eqnarray}

Conservation equation satisfied by the dark energy model NVMCG is

\begin{eqnarray}
\dot{\rho}+3H(\rho+p)=0
\end{eqnarray}

where $H=\frac{\dot{R}}{R}$ is the Hubble parameter.\\

Expression for the energy density for NVMCG model is obtained from
(12) and (15) as \cite{UDebnath}

\begin{eqnarray}
\rho=R^{-3}\exp \left({\frac{3A_0 R^{-n}}{n}}\right) \left[C_0
+\frac{B_0}{A_0} X^{\frac{3(1+\alpha)+n-m}{n}} \Gamma
\left(\frac{m-3(1+\alpha)}{n}, X
R^{-n}\right)\right]^{\frac{1}{1+\alpha}}
\end{eqnarray}

where $C_0$ is an integration constant, $\Gamma (s,t)$ is the
upper incomplete gamma function and $X=\frac{3A_0
(1+\alpha)}{n}$.\\

Following \cite{Michel}, the critical values for this model are as
follows
\begin{eqnarray}
c_{s*}^2=A'+\alpha \frac{B}{\rho_{*} ^{\alpha+1}}\nonumber\\
u^2_*=\frac{A' \rho_*^{\alpha+1}+\alpha B}{\rho_*^{\alpha+1}(1+3 A')+3 \alpha B}\nonumber\\
x_*=\frac{\rho_*^{\alpha+1} (1+3 A')+3 \alpha B}{2\left[A' \rho_*^{\alpha+1}+\alpha B\right]}
\end{eqnarray}

Also from equation (6) and (4), we obtain the ratio of number
density of NVMCG near the horizon and at the infinity as in the
following

\begin{eqnarray}
\frac{n(\rho)}{n_\infty}=\left[\frac{B'
-\rho^{\alpha+1}}{B'-\rho_\infty^{\alpha+1}}\right]^{\mu}
\end{eqnarray}

where $\mu=\frac{1}{(1+\alpha)(1+A')}$, $B'=\frac{B}{1+A'}$ and
\begin{eqnarray}
A=-\frac{1}{4}\left[\frac{\rho_*^{\alpha+1}(1+3 A')+3 \alpha B}{A'
\rho_*^{\alpha+1}+\alpha B}\right]^{3/2} \left[\frac{B'
-\rho_{*}^{\alpha+1}}{B'-\rho_\infty^{\alpha+1}}\right]^{\mu}
\end{eqnarray}

Dark energy density $\rho_H$ at the event horizon of the black
hole satisfies the following equation
\begin{eqnarray}
\left[\frac{\rho_H^{\alpha+1}-B'}{\rho_\infty^{\alpha+1}-B'}\right]^{2
\mu-1}=\frac{A}{4}
\left(\frac{\rho_\infty}{\rho_H}\right)^{\alpha}
\end{eqnarray}

Also the energy density can be expressed in terms of $x$ as

\begin{eqnarray}
\left[\left(1-\frac{2}{x}\right)+\frac{A^2}{x^4}\left\{\frac{B'-\rho_\infty^{\alpha+1}}
{B'-\rho^{\alpha+1}}\right\}^{\mu}\right]
\left(\frac{\rho_\infty}{\rho}\right)^{2 \alpha} \left[\frac{B'
-\rho^{\alpha+1}}{B'-\rho_\infty^{\alpha+1}}\right]^{2 (1-\mu)}=1
\end{eqnarray}

Accreted fluid velocity with respect to radial coordinate $x$ for
NVMCG onto Schwarzschild black hole are drawn in figure 1. The
fluid velocity decreases as $x$ increases, i.e., accreted fluid
velocity is high near the black hole and velocity is low when the
fluid is far from the black hole. The black hole mass increases
for the NVMCG dark energy type accreted fluid as time goes on,
which is shown in figure 3. Also relative density
$\rho/\rho_{\infty}$ of accreted fluid increases when $x$ increases
from the black hole (figure 5).\\

\begin{figure}
\includegraphics[height=1.7in]{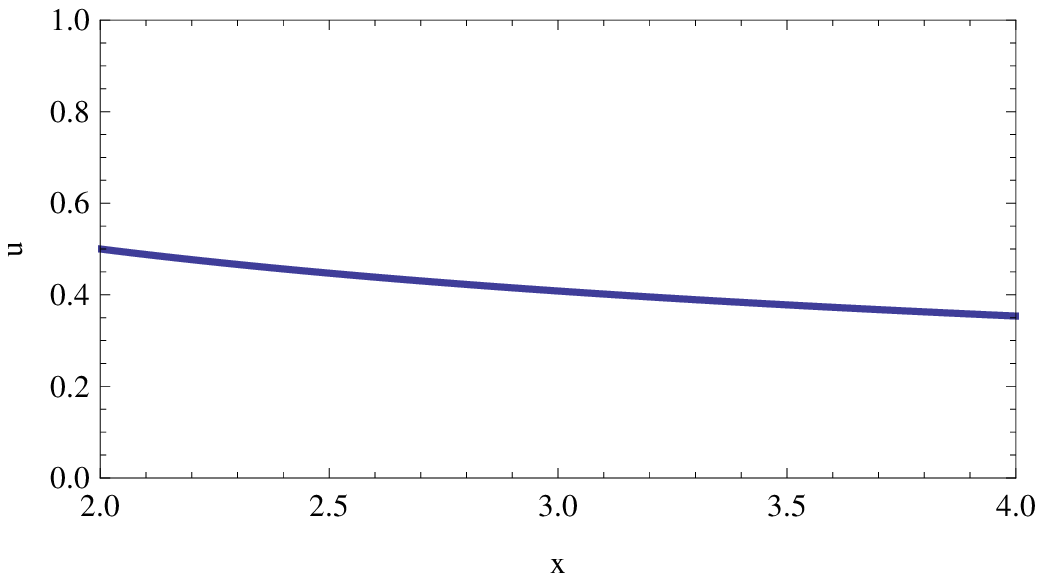}~~~\epsfxsize = 3 in \epsfysize =1.7 in
\epsfbox{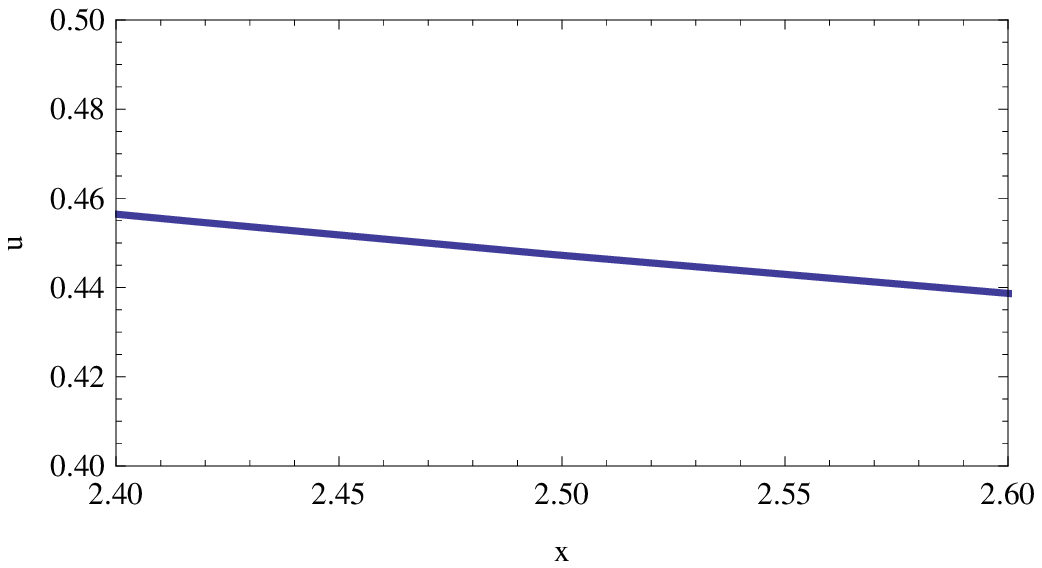}\\
~~FIG.1~~~~~~~~~~~~~~~~~~~~~~~~~~~~~~~~~~~~~~~~~~~~~~~~~~~~~~~~~~~~~~~~~~~~FIG.2\\
\caption{Accreted fluid velocity with respect to radial coordinate
$x$ for NVMCG onto Schwarzschild black hole.} \caption{Accreted
fluid velocity with respect to radial coordinate $x$ for GCCG onto
Schwarzschild black hole.}
\end{figure}

\subsection{Model II: Generalized Cosmic Chaplygin Gas as the dark energy Model}

A new version of Chaplygin gas which is known as Generalized
Cosmic Chaplygin Gas (GCCG) \cite{Diaz,Chak} obeys the equation of
state

\begin{eqnarray}
p=-\rho^{-\alpha}\left[C+\left(\rho^{1+\alpha}-C\right)^{-w}\right]
\end{eqnarray}

where $C=\frac{A''}{(1+w)}-1$, $A''$ takes either positive or
negative constant, $-l<w<0$ and $l>1$. The EOS reduces to that of
current Chaplygin unified models for dark matter and dark energy
in the limit $w\rightarrow 0$ and satisfies the conditions: (i) it
becomes a de Sitter fluid at late time and when $w=-1$, (ii) it
reduces to $p=w \rho$ in the limit that the Chaplygin parameter
$A''\rightarrow 0$, (iii) it also reduces to the EOS of current
Chaplygin unified dark matter models at high energy density and
(iv) the evolution of density perturbations derived from the
chosen EOS becomes free from the pathological behaviour of the
matter power spectrum for physically reasonable values of the
involved parameters at late time. This EOS shows dust era in
the past and $\Lambda$CDM in the future.\\

From the conservation equation we have the expression for energy
density in the form \cite{Chak}

\begin{eqnarray}
\rho=\left[C+\left(1+\frac{B}{R^{3(1+\alpha)(1+w)}}\right)^{\frac{1}{1+w}}\right]^{\frac{1}{1+\alpha}}
\end{eqnarray}

Following \cite{Michel}, the critical values are obtained in the
form

\begin{eqnarray}
c_{s*}^2=\rho_*^{-1-\alpha}y\nonumber\\
u_*^2=\frac{y}{\rho_*^{1+\alpha}+3  y}\nonumber\\
x_*=\frac{\rho_*^{1+\alpha}+3 y}{2y}
\end{eqnarray}
where $y=\left[C \alpha+\left(\rho_*^{1+\alpha}-C\right)^{-1-w}
\left\{\rho_*^{1+\alpha}(\alpha+w+\alpha
w)-C\alpha\right\}\right]$.\\

From equations (6) and (4), we obtain the ratio of number density
of GCCG near the horizon and at the infinity as in the following

\begin{eqnarray}
\frac{n(\rho)}{n_\infty}=\left(\frac{ f_{1}(\rho)f_{2}(\rho)}{
f_{1}(\rho_{\infty})f_{2}(\rho_{\infty})}\right)^{\nu}
\end{eqnarray}
where,
\begin{eqnarray}
f_{1}(\rho)=\left[\rho^{-\alpha}\left\{C+(-C+\rho^{1+\alpha})^{-w}\right\}
-\rho\right]~,~f_{2}(\rho)=\rho^{\alpha(1+w)}\left(C
\rho^{-\alpha} -\rho\right)^w~and~\nu=\frac{1}{(1+\alpha)(1+w)}
\end{eqnarray}

Energy density can be expressed in terms of $x$ as

\begin{eqnarray}
\left(1-\frac{2}{x}\right)+\frac{A^2}{x^4}\frac{n_\infty}{n(\rho)}=
\left[\frac{f_{1}(\rho)}{f_{1}(\rho_{\infty})}\right]^{2 (\nu-1)}
\left[\frac{f_{2}(\rho)}{f_{2}(\rho_{\infty})}\right]^{2\nu}
\end{eqnarray}
where $A=-\frac{1}{4}\left(\frac{\rho_*^{1+\alpha}+3 y}{2
y}\right)^{3/2}
 \frac{n(\rho_*)}{n_\infty}$.\\

\begin{figure}[!h]
\epsfxsize = 3.0 in \epsfysize = 3 in
\epsfbox{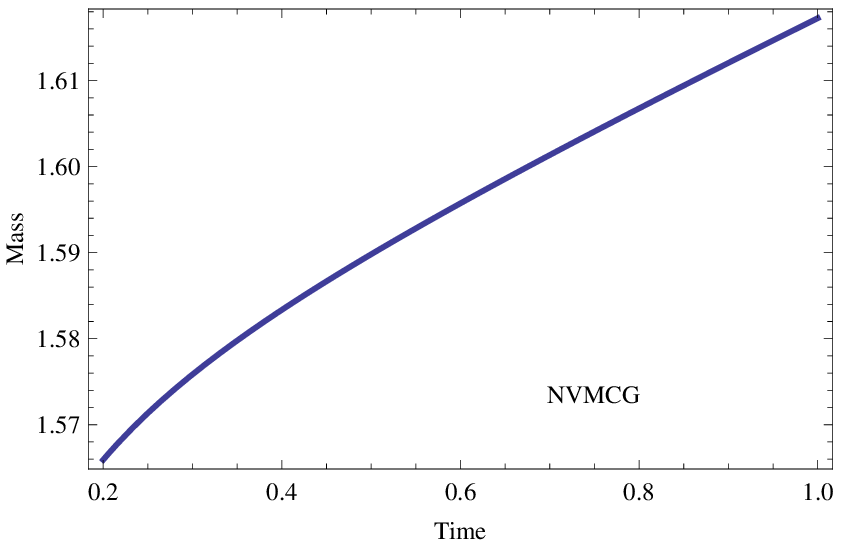}~~~\epsfxsize = 3 in \epsfysize =3 in
\epsfbox{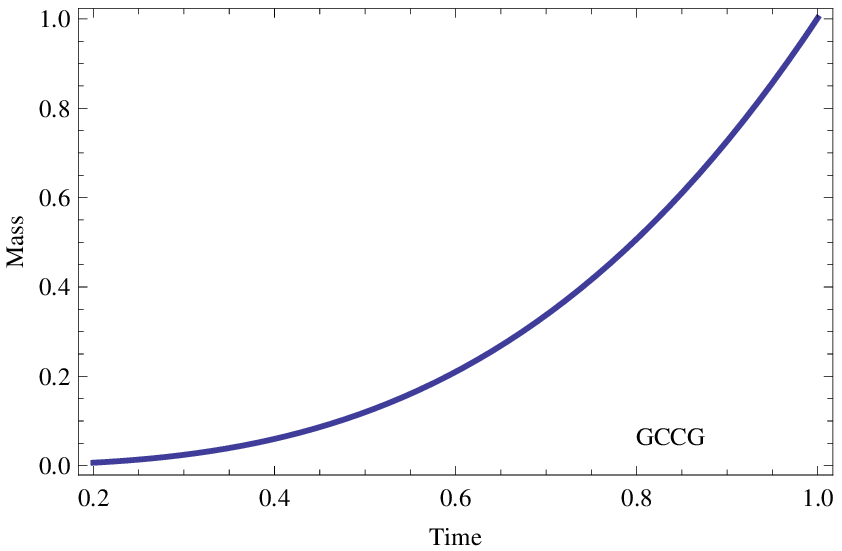}\\
~~FIG.3~~~~~~~~~~~~~~~~~~~~~~~~~~~~~~~~~~~~~~~~~~~~~~~~~~~~~~~~~~~~~~~~~~~~FIG.4\\
\caption{Changes of the mass with respect to time of  NVMCG onto
Schwarzschild black hole.} \caption{Changes of the mass with
respect to time of GCCG onto Schwarzschild black hole. \vspace{.5in}}

\end{figure}

\begin{figure}[!h]
\epsfbox{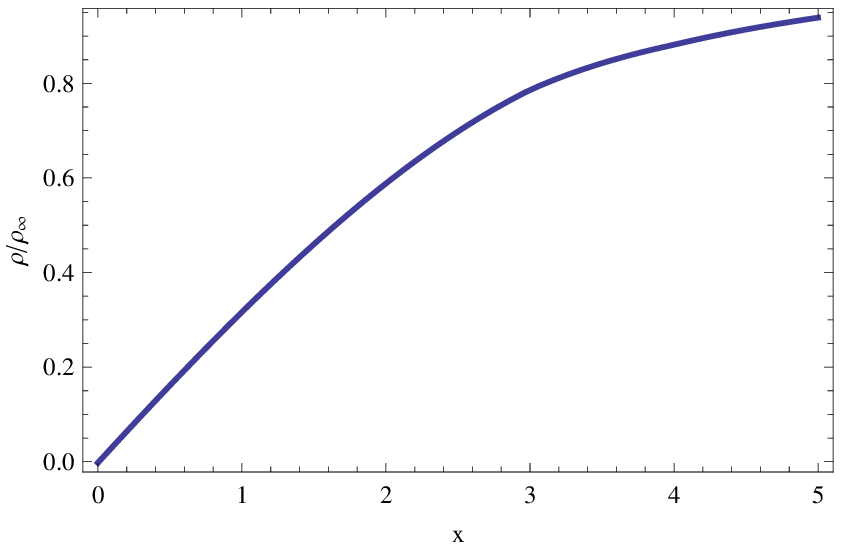}~~~
\includegraphics[height=2.3in]{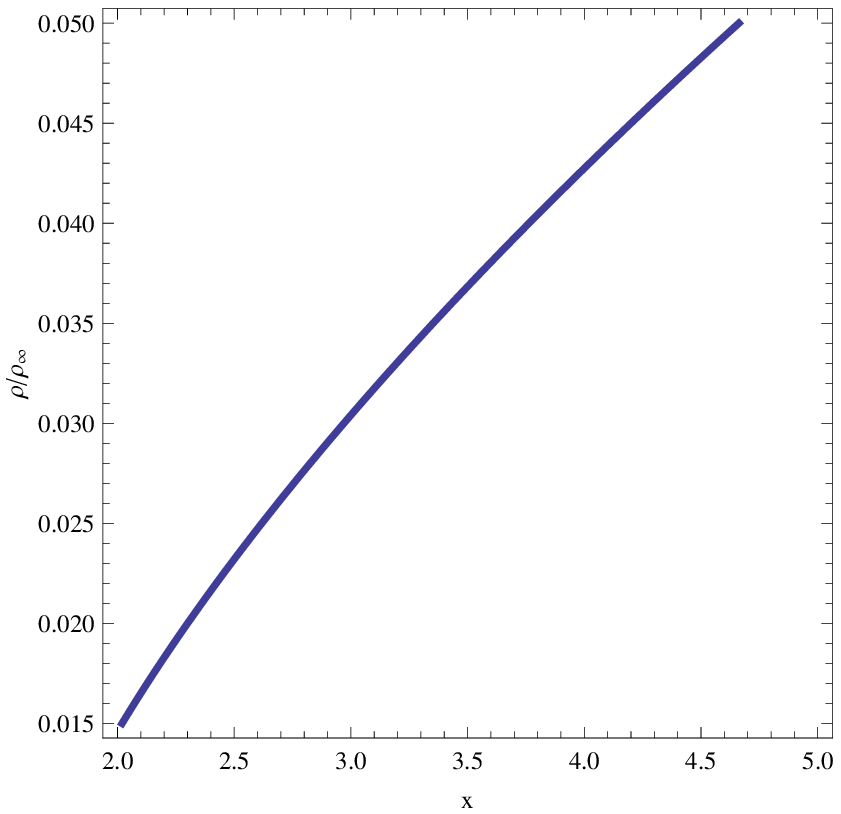}\\
~~FIG.5~~~~~~~~~~~~~~~~~~~~~~~~~~~~~~~~~~~~~~~~~~~~~~~~~~~~~~~~~~~~~~~~~~~~FIG.6\\
\caption{Relative density of accreted fluid with respect to radial
coordinate $x$ for NVMCG onto Schwarzschild black hole.}
\caption{Relative density of accreted fluid with respect to radial
coordinate $x$ for GCCG onto Schwarzschild black hole.}
\end{figure}

Accreted fluid velocity with respect to radial coordinate $x$ for
GCCG onto Schwarzschild black hole are drawn in figure 2. The
fluid velocity decreases as $x$ increases, i.e., accreted fluid
velocity is high near the black hole and velocity is low when the
fluid is far from the black hole. The black hole mass increases
for the GCCG dark energy type accreted fluid as time goes on,
which is shown in figure 4. Also relative density
$\rho/\rho_{\infty}$ of accreted fluid increases when $x$
increases from the black hole (figure 6).\\

\section{General accretion of dark energy onto Kerr-Newman Black hole}

For the rotating, charged, stationary, axisymmetric black hole,
let us consider Kerr-Newman space-time metric (considering
$G=c=1$) \cite{Pedro} prescribed by the line element

\begin{eqnarray}
ds^2=\left(1+\frac{Q^2-2Mr}{r^2+a^2 \cos^2 \theta}\right) dt^2+\frac{2a(2Mr-Q^2)\sin^2
\theta}{r^2+a^2 \cos^2 \theta} dt d\phi-\frac{r^2+a^2 \cos^2 \theta}{r^2+a^2 +Q^2-2Mr} dr^2\nonumber\\
 -\left(r^2+a^2 \cos^2 \theta\right) d\theta^2-\left[r^2+a^2+\frac{\left(2Mr-Q^2\right)a^2
 \sin^2 \theta}{r^2+a^2 \cos^2 \theta}\right]\sin^2 \theta d\phi^2
\end{eqnarray}
where $M$ and $Q$ are respectively mass, electric charges of the
black hole. Also $a=J/M$ is the specific angular momentum per unit
mass and $J$ is the total angular momentum of the black hole.
Since Kerr-Newman metric is static, the time evolution induced by
accretion will be taken into account by the time dependence of the
scale factor entering the integrated conservation laws and the
rate equations for mass and angular momentum. Also the energy
tensor $T_{\mu\nu}$ for dark energy satisfies the relation (2).
The first integral of motion comes from the time component of the
energy-momentum conservation law $T^{\mu\nu}_{;\nu}=0$ which can
be put in the following form for Kerr-Newman metric:

\begin{eqnarray}
\frac{d}{dr}\left[(p+\rho) \left(1+\frac{Q^2-2Mr}{r^2+a^2 \cos^2 \theta}\right)\frac{dt}{ds}\frac{dr}{ds} \right]
+\frac{2r}{r^2+a^2 \cos^2 \theta} (p+\rho) \left(1+\frac{Q^2-2Mr}{r^2+a^2 \cos^2 \theta}\right)\frac{dt}{ds}\frac{dr}{ds}\nonumber\\
+\frac{d}{d\theta} \left[(p+\rho) \left(1+\frac{Q^2-2Mr}{r^2+a^2 \cos^2 \theta}\right)\frac{dt}{ds}\frac{d\theta}{ds})\right]
+\left[\left(\frac{\cos \theta}{\sin\theta}-\frac{2 a^2 \sin \theta \cos\theta}{r^2+a^2 \cos^2 \theta}\right)(p+\rho)
\left(1+\frac{Q^2-2Mr}{r^2+a^2 \cos^2 \theta}\right)\right]=0\nonumber\\
\end{eqnarray}

In general, $\rho$ and $p$ are functions of $t$, $r$ and $\theta$,
but in our considered dark energy models $\rho$ and $p$ depend
only on time $t$. Keeping $\theta $ constant and radial four
velocity component $u=\frac{dr}{ds}$ we get the expression for
rate of change of mass due to accretion of dark energy
\cite{Pedro}
\begin{eqnarray}
\dot{M}=-\int T_0^r dS=\frac{4 \pi A_M M^3 r}{J} \arctan \left(\frac{J}{Mr}\right) (p+\rho)
\end{eqnarray}

where $dS=r^2 \sin \theta d\theta d\phi$ with $r$ and $J$ are
constants. Which gives the expression

\begin{eqnarray}
I_M=\int_{M_0}^{M}\frac{J dM}{M^3 r \arctan \left(\frac{J}{Mr}\right)}=4 \pi A_M \int_{t_0}^{t} (p+\rho) dt
\end{eqnarray}
with
\begin{eqnarray}
A_M=-\frac{u}{M^2} (r^2+a^2 \cos^2 \theta) \exp \left[\int^{\rho}_{\rho_\infty}\frac{d \rho}{\rho+p(\rho)}\right]
\end{eqnarray}

When a fluid flow directed toward the black hole we have $u<0$ and $A_M>0$ is dimensionless constant.\\

Again keeping $r$ as constant and we get the rate of change of angular momentum of Kerr-Newman black hole as \cite{Pedro}
\begin{eqnarray}
\dot{a}=-\int rT_0^\theta dS=\frac{2 \pi^2 A_a ar^2 (p+\rho)}{\sqrt{r^2+a^2}}
\end{eqnarray}

where $dS=r^2 \sin \theta d\theta d\phi$, $\theta$ constant. Equation (33) simplifies to

\begin{eqnarray}
I_a=\int_{a_0}^{a} \frac{\sqrt{r^2}+a^2}{a r^2}da=2 \pi^2 A_a\int_{t_0}^{t} (p+\rho) dt
\end{eqnarray}

with
\begin{eqnarray}
A_a=-\frac{1}{a} \omega \sin \theta (r^2+a^2 \cos^2 \theta) \exp
\left[\int^{\rho}_{\rho_\infty}\frac{d
\rho}{\rho+p(\rho)}\right]~~,~~\theta = constant.
\end{eqnarray}

When a fluid flow is directed toward the black hole then $\omega=\frac{d\theta}{ds}<0$  for  $A_a>0$.\\

Expression for the rate of change of total angular momentum, with
$M$, $r$ constant, is given by \cite{Pedro}
\begin{eqnarray}
\dot{J}=-\int \left(M r T_0^\theta+ aT_0^r\right) dS=\pi (p+\rho) \left[\frac{2J\pi A_a r}{\sqrt{1+\frac{J^2}{M^2 r^2}}}+4 A_M M^2 r \arctan \left(\frac{J}{Mr}\right)\right]
\end{eqnarray}

Which reduces to
\begin{eqnarray}
I_J=\int_{J_0}^{J} \frac{dJ}{\left[\frac{2J\pi A_a r}{\sqrt{1+\frac{J^2}{M^2 r^2}}}+4 A_M M^2 r \arctan \left(\frac{J}{Mr}\right)\right]}=\pi \int_{t_0}^{t} (p+\rho) dt
\end{eqnarray}

We shall obtain the quantity $\dot{M}, \dot{a}, \dot{J}$ for the
following cosmological models:

\subsection{Model I: New Variable Modified Chaplygin Gas as dark
energy model}

For the NVMCG  satisfying the equations (11-15) we derive the
following quantities

\begin{eqnarray}
\sqrt{\frac{3}{8 \pi A_M^2}} I_M=\sqrt{\rho_0}-\sqrt{\rho}
\end{eqnarray}

\begin{eqnarray}
\sqrt{\frac{3}{2 \pi^3 A_a^2}} I_a=\sqrt{\rho_0}-\sqrt{\rho}
\end{eqnarray}

\begin{eqnarray}
\sqrt{\frac{6}{\pi}} I_J=\sqrt{\rho_0}-\sqrt{\rho}
\end{eqnarray}

where $\rho$ satisfying the relation (16) and from this equation
we obtain the present value of density $\rho_0$ which is given by

\begin{eqnarray}
\rho_0=R_0^{-3}\exp \left({\frac{3A_0 R_0^{-n}}{n}}\right) \left[C_0 +\frac{B_0}{A_0}
\left(\frac{3A_0 (1+\alpha)}{n}\right)^{\frac{3(1+\alpha)+n-m}{n}} \Gamma
\left(\frac{m-3(1+\alpha)}{n}, \frac{3A_0(1+\alpha)}{n}R_0^{-n}\right)\right]^{\frac{1}{1+\alpha}}
\end{eqnarray}

where $R_0$ is the present value of scale factor. Changes of the
mass with respect to time of  NVMCG onto Kerr-Newman black hole
with the constant $J$ and variable $J$ are drawn in figures 7 and
8 and  they are increasing. The angular momentum and the total
angular momentum with time are shown in figures 9 and 10 and they are increasing with time.\\

\begin{figure}[!h]
\epsfxsize = 3.0 in \epsfysize = 3 in
\epsfbox{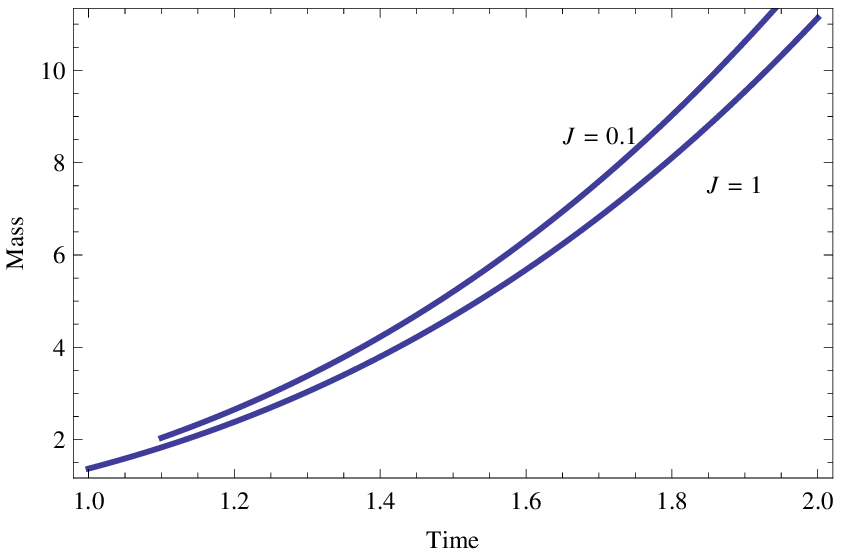}~~~\epsfxsize = 3 in \epsfysize =3 in
\epsfbox{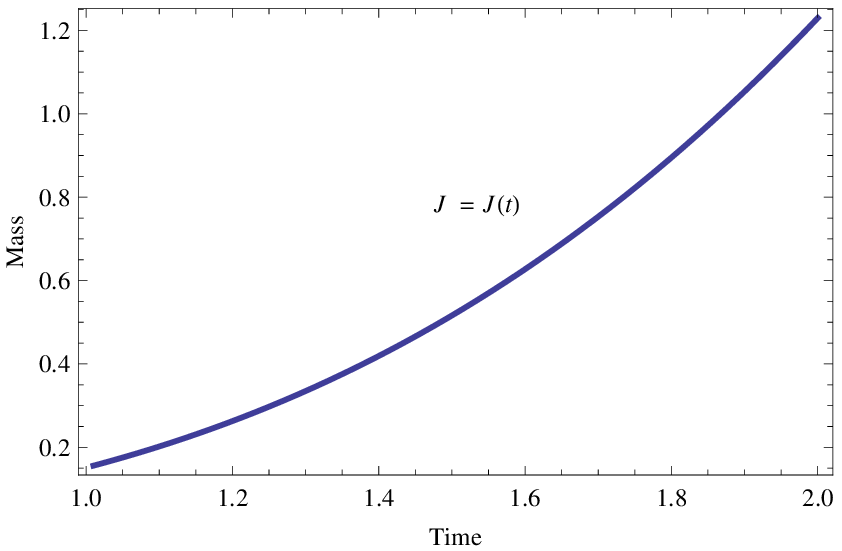}\\
~~FIG.7~~~~~~~~~~~~~~~~~~~~~~~~~~~~~~~~~~~~~~~~~~~~~~~~~~~~~~~~~~~~~~~~~~~~FIG.8\\
\caption{Changes of the mass with respect to time of  NVMCG onto
Kerr-Newman black hole with the constant $J$.} \caption{Changes of
the mass with respect to time of NVMCG onto Kerr-Newman black hole
with $J=J(t)$.}
\end{figure}

\begin{figure}[!h]
\epsfxsize = 3.0 in \epsfysize = 3 in
\epsfbox{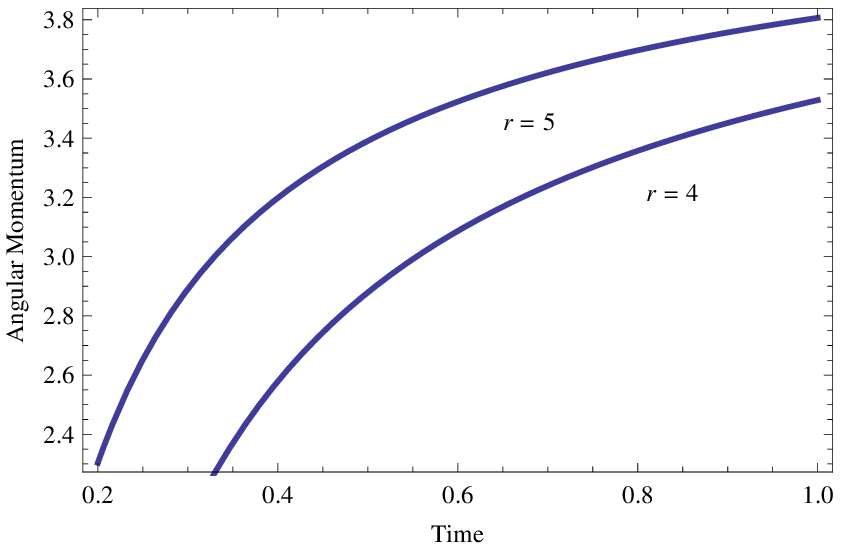}~~~\epsfxsize = 3 in \epsfysize =3 in
\epsfbox{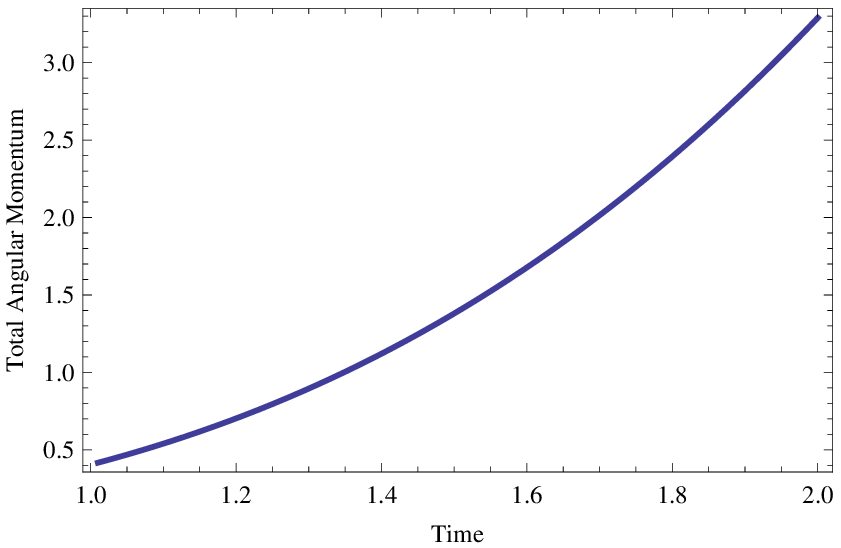}
~~FIG.9~~~~~~~~~~~~~~~~~~~~~~~~~~~~~~~~~~~~~~~~~~~~~~~~~~~~~~~~~~~~~~~~~~~~FIG.10\\
\caption{Changes of the specific angular momentum ($a$) with respect to time of
NVMCG onto Kerr-Newman black hole with the constant $r$.}
\caption{Changes of the total angular momentum ($J$) with respect to
time of NVMCG onto Kerr-Newman black hole.}
\end{figure}

\subsection{Model II: Generalized Cosmic Chaplygin Gas as the
dark energy Model}

For the Generalized Cosmic Chaplygin Gas (GCCG)  satisfying the
equations (12-14) and (21,22) we derive

\begin{eqnarray}
R^{3(1+\alpha)(1+w)}=\frac{B}{1-\left[\left(\sqrt{\rho_0}-\sqrt{\frac{3}{8 \pi A_M^2}} I_M\right)^{2(1+\alpha)}-C\right]^{(1+w)}}
\end{eqnarray}

\begin{eqnarray}
R^{3(1+\alpha)(1+w)}=\frac{B}{1-\left[\left(\sqrt{\rho_0}-\sqrt{\frac{3}{2 \pi^3 A_a^2}} I_a\right)^{2(1+\alpha)}-C\right]^{(1+w)}}
\end{eqnarray}

\begin{eqnarray}
R^{3(1+\alpha)(1+w)}=\frac{B}{1-\left[\left(\sqrt{\rho_0}-\sqrt{\frac{6}{ \pi }} I_J\right)^{2(1+\alpha)}-C\right]^{(1+w)}}
\end{eqnarray}

with,
\begin{eqnarray}
\rho_0=\left[C+\left(1+\frac{B}{R_0^{3(1+\alpha)(1+w)}}\right)^{\frac{1}{1+w}}\right]^{\frac{1}{1+\alpha}}
\end{eqnarray}

\begin{figure}[!h]
\epsfxsize = 3.0 in \epsfysize = 3 in
\epsfbox{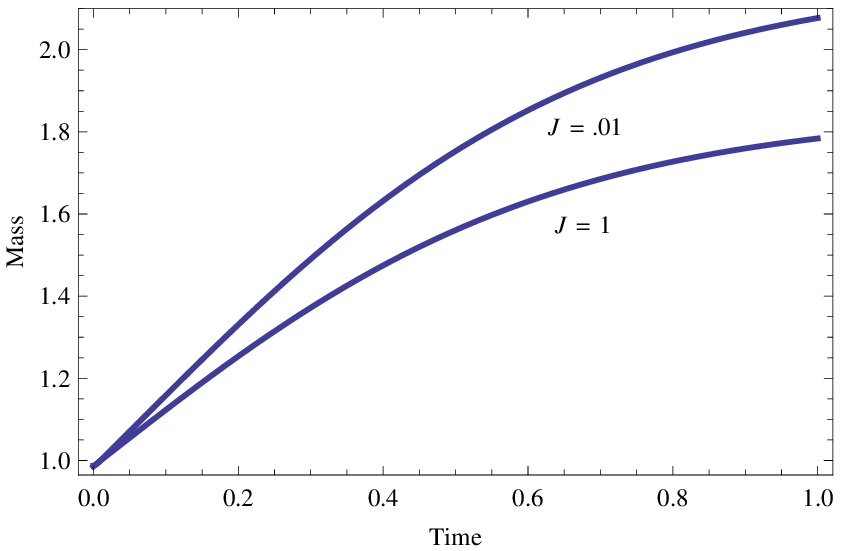}~~~\epsfxsize = 3 in \epsfysize =3 in
\epsfbox{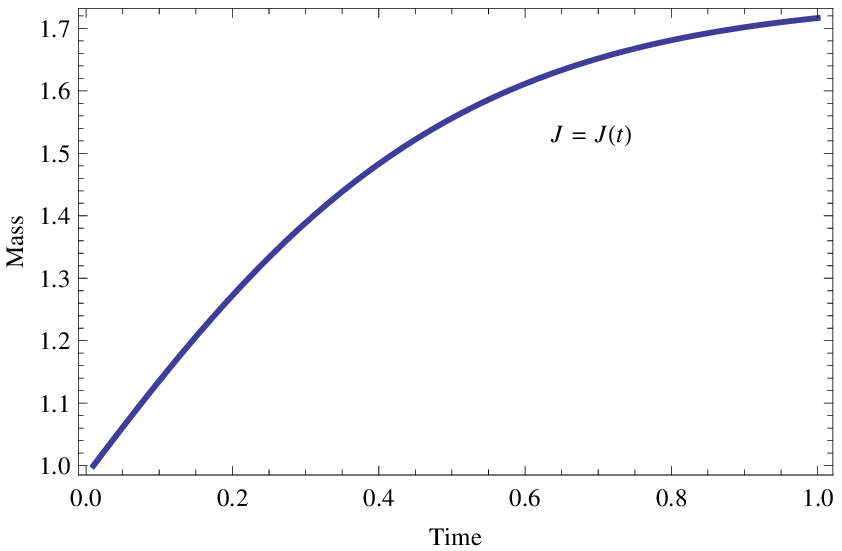}\\
~~FIG.11~~~~~~~~~~~~~~~~~~~~~~~~~~~~~~~~~~~~~~~~~~~~~~~~~~~~~~~~~~~~~~~~~~~~FIG.12\\
\caption{Changes of the mass with respect to time of  GCCG onto
Kerr-Newman black hole with the constant $J$.} \caption{Changes of
the mass with respect to time of GCCG onto Kerr-Newman black hole
with $J=J(t)$.}
\end{figure}

\begin{figure}[!h]
\epsfxsize = 3.0 in \epsfysize = 3 in
\epsfbox{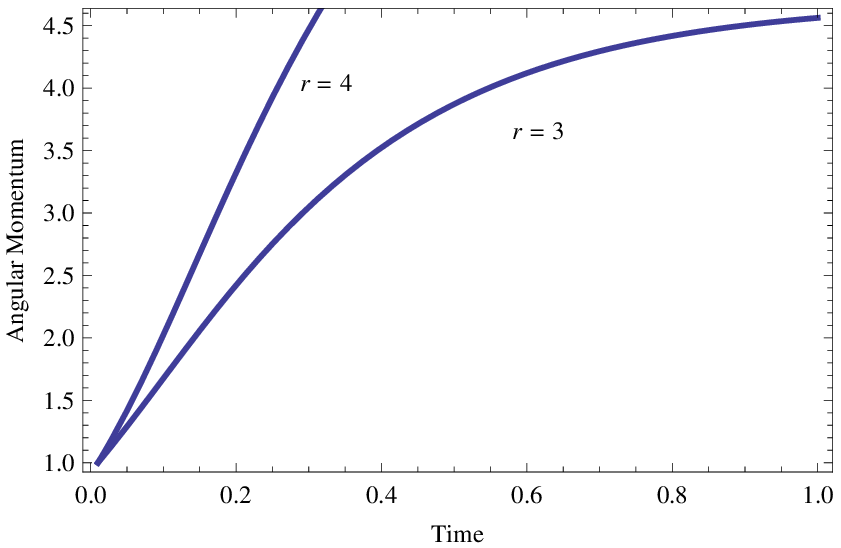}~~~\epsfxsize = 3 in \epsfysize =3 in
\epsfbox{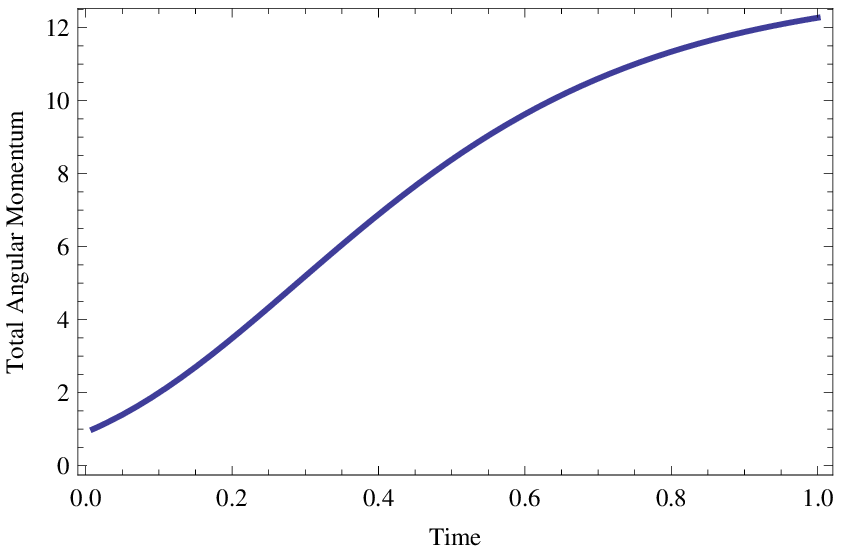}
~~FIG.13~~~~~~~~~~~~~~~~~~~~~~~~~~~~~~~~~~~~~~~~~~~~~~~~~~~~~~~~~~~~~~~~~~~~FIG.14\\
\caption{Changes of the specific angular momentum ($a$) with respect to time of
GCCG onto Kerr-Newman black hole with the constant $r$.}
\caption{Changes of the total angular momentum ($J$) with respect to
time of GCCG onto Kerr-Newman black hole.}
\end{figure}

Changes of the mass with respect to time of  GCCG onto Kerr-Newman
black hole with the constant $J$ and variable $J$ are drawn in
figures 11 and 12 and  they are increasing. The angular momentum
and the total angular momentum with time are shown in figures 13
and 14 and they are increasing with time.

\section{Discussions}

In this work, we have studied recently proposed two types of dark
energy models like new variable modified Chaplygin gas (NVMCG) and
generalized cosmic Chaplygin gas (GCCG). Accretion of NVMCG and
GCCG onto the Schwarzschild and Kerr-Newman Black holes have been
discussed. Our dark energy fluids violate the strong energy
condition ($\rho+3p<0$ in late epoch), but do not violate the weak
energy condition ($\rho+p>0$). So the models drive only
quintessence scenario in late epoch, but do not generate the
phantom epoch (in our choice). We find the expression of the
critical four velocity component which gradually decreases for the
fluid flow towards the Schwarzschild as well as Kerr-Newman Black
holes. Astrophysically, mass of the black hole is a dynamical
quantity, so the nature of the mass function is important in our
black hole models for different dark energy filled universe.
Previously Babichev et al \cite{Babichev} have shown that the mass
of black hole decreases due to phantom energy accretion. We here
found the expression for change of masses of the Schwarzschild and
Kerr-Newman black holes in both the dark energy models and have
seen that they are increasing in course of time. Since our
considered dark energy candidates do not violate weak energy
condition, so the dynamical mass of the black holes could not
decaying by the accretion of dark energies, though the pressures
of the dark energies are outside the black holes. The relative
density $\rho/\rho_{\infty}$ increases as $r$ increases outside
the black hole. For the most generalized Kerr-Newman black hole
(which is rotating and charged) we have obtained the specific
angular momentum ($a$) and total angular momentum ($J$). We showed
that in both cases due to accretion of the dark energy mass of the
black hole increases and angular momentum increases in case of
Kerr-Newman black hole. The mass of the black hole increases for
constant and variable $J$. The relative density shows the same
nature as well as Schwarzschild black hole. \\\\

{\bf Acknowledgement:}\\

One of the authors (JB) is thankful to CSIR, Govt of India for
providing Junior Research Fellowship. The authors are thankful to
IUCAA, Pune, India for warm hospitality where part of the work was
carried out.\\

\end{document}